\documentclass{JINST}
\usepackage{amsmath}
\usepackage{txfonts}
\usepackage{latexsym}
\usepackage{afterpage}
\usepackage[numbers]{natbib}

\newcommand{\Planck}{\textsc{Planck}}

\newcommand{\unit}[1]{\text{#1}}

\title{Dynamic Validation of the \Planck/LFI Thermal Model}

\author{M.~Tomasi, B.~Cappellini, A.~Gregorio, F.~Colombo, M.~Lapolla, L.~Terenzi, G.~Morgante, M.~Bersanelli, R.~C.~Butler, S.~Galeotta, N.~Mandolesi, M.~Maris, A.~Mennella, L.~Valenziano, A.~Zacchei.}

\abstract{The Low Frequency Instrument (LFI) is an array of cryogenically cooled radiometers on board the \Planck{} satellite, designed to measure the temperature and polarization anisotropies of the cosmic microwave backgrond (CMB) at 30, 44 and 70~GHz. The thermal requirements of the LFI, and in particular the stringent limits to acceptable thermal fluctuations in the 20\,K focal plane, are a critical element to achieve the instrument scientific performance. Thermal tests were carried out as part of the on-ground calibration campaign at various stages of instrument integration. In this paper we describe the results and analysis of the tests on the LFI flight model (FM) performed at Thales Laboratories in Milan (Italy) during 2006, with the purpose of experimentally sampling the thermal transfer functions and consequently validating the numerical thermal model describing the dynamic response of the LFI focal plane. This model has been used extensively to assess the ability of LFI to achieve its scientific goals: its validation is therefore extremely important in the context of the \Planck{} mission. Our analysis shows that the measured thermal properties of the instrument show a thermal damping level better than predicted, therefore further reducing the expected systematic effect induced in the LFI maps. We then propose an explanation of the increased damping in terms of non-ideal thermal contacts.}

\keywords{Cosmic microwave background - Instrumentation: detectors - Methods: data analysis - Methods: numerical - Thermal modeling - Temperature fluctuations - Heat conduction - Contact resistance}

\begin{document}


\section{Introduction}

\Planck{} is a space mission of the European Space Agency (ESA) whose main objective is to image the temperature and polarization anisotropies of the Cosmic Microwave Background (CMB) with unprecedented sensitivity, angular resolution, sky coverage and frequency coverage \cite{2009_COM_Mission}. Two instruments share the focal plane of the \Planck{} 1.5 m telescope: HFI (High Frequency Instrument, \cite{2009_HFI_Instrument}), an array of 52 bolometers; and LFI (Low Frequency Instrument, \cite{2009_LFI_cal_M2}), an array of 22 pseudo-correlation radiometers. Both HFI and LFI need to be cooled to cryogenic temperatures (0.1\,mK for the HFI bolometers and 20\,K for the LFI radiometers) in order to meet their scientific requirements.

\begin{figure}
  \begin{center}
    \begin{tabular}{cc}
      \includegraphics[width=0.35\textwidth]{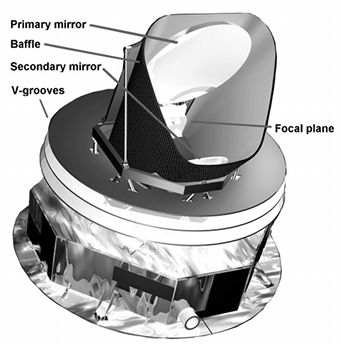} &
      \includegraphics[width=0.35\textwidth]{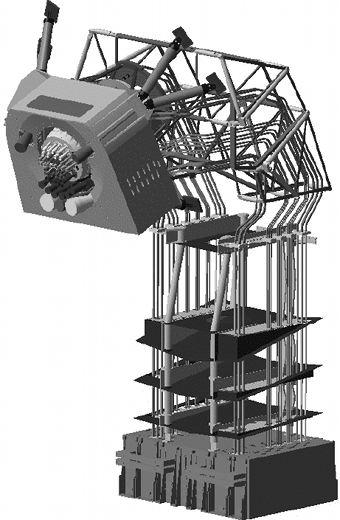} \\
    \end{tabular}
  \end{center}
  \caption{\label{fig:planck} \textbf{Left}: schematics of the \Planck{} satellite. The warm service module ($\sim$ 300\,K) is thermally decoupled from the focal plane and the telescope by means of three conical radiators. The LFI focal plane is cooled to 20\,K by a hydrogen sorption cooler which also acts as a pre-cooling stage for the HFI 4\,K cooler that cools the HFI focal plane and the LFI reference load system \cite{2009_LFI_cal_R1}. \textbf{Right}: detailed view of the LFI structure. On top, the cold Focal Plane Unit (FPU) with both the LFI and HFI feed horns is shown. A set of waveguides connect the FPU with the warm (300\,K) Back End Unit (BEU), shown at the bottom.}
\end{figure}

\Planck{} is the first space mission devoted to the measurement of CMB anisotropies that uses cryogenic instruments, and it presents extreme challenges in its thermal design. The thermal system uses a mixed passive/active approach to reach the required working temperatures. Three thermal shields radiate heat and decouple the warm service module ($\sim$ 300\,K) from the telescope, passively cooled to $<$50\,K, and the focal instruments. Active cooling is provided by three cryocoolers: a hydrogen sorption cooler (18-20\,K), a Stirling cooler (4\,K) and a $^3$He-$^4$He diluition cooler (0.1\,K).

Within such a complex cryogenic system, thermal effects are expected to be the most important source of systematics. Temperature instabilities in the detectors can degrade the scientific output of \Planck{}, which must be able to measure fluctuations in the CMB signal with a sensitivity of $\delta T/T \sim 10^{-6}$ per pixel on the final maps. For this reason, one of the main driving requirements in the design of \Planck{} has been to maximize the damping of thermal fluctuations propagating through the two instruments, and an extensive set of thermal tests has been performed at both instrument and satellite level in order to verify that the thermal stability requirements are satisfied.

The LFI focal plane, cooled to 20\,K, includes front end passive components (corrugated feed horns and orthomode transducers), hybrid couplers, and state-of-the-art HEMT (High Electron Mobility Transistor) low noise cryogenic amplifiers providing $\sim$ 30\,dB of amplification. The thermal susceptibility of the front end components, in particular the thermal coupling to the gain of the HEMT amplifiers and to the insertion loss of the front-end passive components, impose very stringent temperature stability requirements ($\sim$ $\mu$K level) on the 20\,K stage.

One of the most important systematic effects in the \Planck{} measurements is the propagation of fluctuations from the 18-20\,K sorption cooler \cite{2009_LFI_cal_M1}, whose stability at the cold end is limited by the thermal cycling of the sorbent compressors \cite{bhandari2004,2009_LFI_Morgante_SCS}. In fact, fluctuations in the LFI 20\,K stage introduce a potentially serious systematic effect in the LFI science as they may mimic brightness changes in the beam as the satellite scans through the sky \cite{2009_LFI_cal_R6}. While the  pseudo-correlation design of the LFI receivers suppress to first order thermal fluctuations \cite{2002MennellaThermal}, residual effects must be damped to extremely low levels (see table~\ref{tab:ErrorBudget}). The combination of radiometer susceptibility, expected fluctuations in the 20\,K stage, and thermal damping from the instrument were the key factors in the instrument design to ensure adequate stability. 

To quantify thermal damping factors in the focal plane, a numerical thermal model was developed by Thales Alenia Space. The model has been used extensively since the design phases (1) to study how thermal systematic effects propagate to the radiometer front ends, and from the results of this analysis (2) to estimate the stability requirements needed for \Planck{}/LFI to meet its scientific objectives.

In this paper we report the measurements of the propagation of fluctuations in the LFI flight model focal plane after the LFI was assembled in the Thales Alenia Space laboratories in Milan (Italy). The first product of our analysis is the sampling of the dynamical thermal transfer function between the focal plane cold end and a set of points on the focal plane itself at three frequencies. We then use such measurements to validate the numerical thermal model of the focal plane, thus confirming (1) the validity of the temperature stability requirements on the \Planck{}/LFI focal plane, which have driven the design of \Planck{}, and (2) the validity of a number of works that have used the estimates of the model \cite{2002MennellaThermal,spieThermalModels,2002A&A...384..736M,2009AshdownLFIMaps}. Finally, we use the measured transfer functions to estimate the improvement in the stability of the receivers over the LFI requirements and provide an explanation of such improvements in terms of non-ideal thermal contacts.

The outline of this article is as follows: Sec.~\ref{sec:Planck} describes the \Planck{}/LFI instrument, the cooling system of \Planck{} and the thermal model used for the characterization of the LFI focal plane. Sec.~\ref{sec:dynamicValidation} describes the extraction of the focal plane thermal transfer functions and the validation of the thermal model of the focal plane. Sec.~\ref{sec:stabilityEstimation} shows how much the measured transfer functions allow to relax the reqmirement on the thermal stability of the radiometers. In Sec.~\ref{sec:contactR}  we propose our explanation for the discrepancies between the model and the measurements (i.e.\ better damping) in terms of contact resistance between the cold end of the focal plane and the focal plane itself. Finally, in Sec.~\ref{sec:conclusions} we report the conclusions of our work.

\section{The \Planck{}/LFI Instrument}
\label{sec:Planck}

\subsection{Overview of \Planck{}/LFI}

The \Planck/LFI instrument \cite{2009_LFI_cal_M2} is an array of 22 pseudo-correlation radiometric receivers centered at 30, 44 and 70\,GHz and cooled to $\sim$ 20\,K by a vibrationless hydrogen sorption cooler \cite{2009_LFI_Morgante_SCS}. Each LFI receiver measures the sky signal ($\sim$ 2.7\,K) by comparing it with the signal coming from a stable reference load ($\sim$ 4.5\,K). This load is thermally connected to the 4\,K HFI shield.




In order to minimise power dissipation in the focal plane, the LFI Radiometric Array Assembly (RAA) is split into two subassemblies (see fig.~\ref{fig:planck}): the Front End Unit (FEU), mounted on the focal plane and actively cooled at 20\,K, and the Back End Unit (BEU) on top of the \Planck{} Service Module at about 300\,K. The FEU is connected to the BEU by means of waveguides (WGs), which carry the microwave signals and by means of the cryo harness that carries the bias currents for the front-end active components. Both the WGs and the cryo harness are connected to three passive radiators (V-grooves) that dissipate the heat coming from the BEU.

\begin{figure}[tf]
  \centering
  \includegraphics[width=0.75\textwidth]{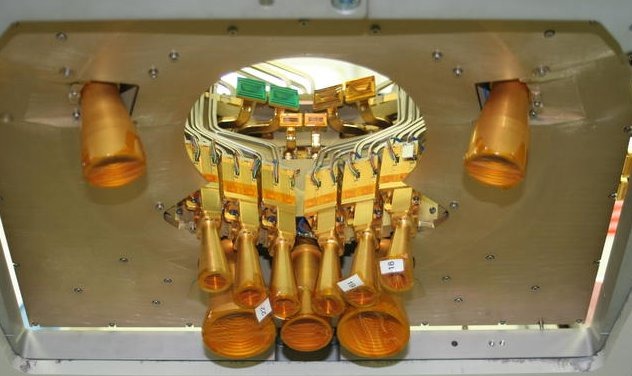}
  \caption{\label{fig:focalPlane} The LFI focal plane during the integration of the flight instrument in the Thales Alenia Space laboratories (Milan, Italy). Note the 11 feed horns (2 centered at 30\,GHz, 3 at 44\,GHz and 6 at 70\,GHz).}
\end{figure}

The goal of carefully controlling thermal systematic effects has driven the design of LFI, which has been optimized in order to maximize the stability to thermal fluctuations, especially those propagating through the focal plane (see fig.~\ref{fig:focalPlane}). In order to improve the focal plane stability, a Temperature Stabilization Assembly (TSA) has been implemented in order to reduce temperature fluctuations at the sorption cooler cold end by one order of magnitude. Such design and optimizations have been made by means of stability estimates (described in Sec.~\ref{sec:stabilityEstimation}) based on a numerical thermal model of the focal plane. The latter is the main topic of this article.

In order to verify that the thermal performance of LFI is compliant with the requirements, the instrument has undergone an extensive thermal analysis and test campaign to evaluate if the measured thermal damping is comparable (or better) to thermal model estimates. This is of capital importance: temperature fluctuations propagating to the amplifiers can induce gain and insertion loss changes that in turns produce systematic variations in the output antenna temperature \cite{2009_LFI_cal_R6}. Therefore, any uncontrolled fluctuation at this level might compromise the whole scientific output of the instrument.

\begin{table}[tf]
  \centering
  \caption{\label{tab:ErrorBudget} Total error budget allocated to thermal fluctuations in the LFI instrument \cite{2009_LFI_cal_M2}. All the numbers are in $\mu$K. The first three columns contain the maximum error for high-frequency fluctuations (h.f.), i.e. $>1/60\,\unit{Hz}$. The last two columns contain the error budget for low-frequency periodic fluctuations and spin-synchronous variations. The BEU (Back-End Unit) comprises the warm part of the radiometers ($\sim$300\,K), the DAE (Digital Acquisition Electronics) is the back-end module devoted to the digitization of the radiometric signal.}
  \begin{tabular}{|cccccc|}
    \hline
    Source & H.f.@30\,GHz & H.f.@44\,GHz & H.f.@70\,GHz & Periodic [$\mu$K] & Spin-s. [$\mu$K] \\
    \hline
    Focal plane & 14.8 & 20.5 & 36.4 & 0.9 & 0.45 \\
    Waveguides  & 14.8 & 20.5 & 36.4 & 0.4 & 0.4 \\
    BEU         & 14.8 & 20.5 & 36.4 & 0.4 & 0.4 \\
    DAE         & 14.8 & 20.5 & 36.4 & 0.4 & 0.4 \\
    \hline
    Total       & 29.6 & 41.0 & 72.8 & 1.1 & 0.8 \\
    \hline
  \end{tabular}
  \label{tab:ThermalErrorBudget}
\end{table}

\subsection{The Thermal Model of the LFI Focal Plane}
\label{sec:thermalModels}

\begin{figure}
  \centering
  \includegraphics[width=\columnwidth]{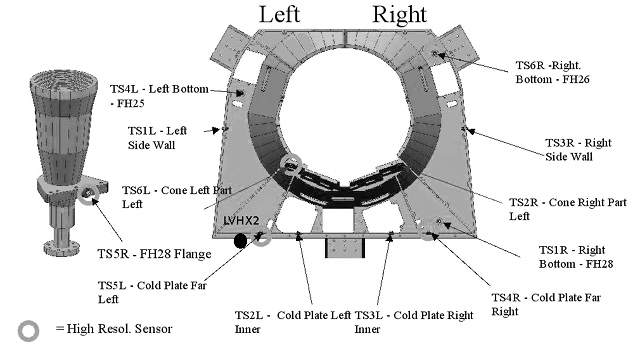}
  \caption{\label{fig:temperatureSensors} Placement of the temperature sensors in the LFI focal plane. One of the sensors (TS5R) has been placed on the flange that holds feed horn \#28 (left). Two sorption coolers are used in \Planck{}; the position of the cold end used during the LFI RAA test campaign (LVHX2) is shown with a black point on the bottom left. All the sensors were included in our analysis, with the exception of TS2L and TS5L (see text).}
\end{figure}

A detailed thermal model of the focal plane was developed during the design of LFI by Thales/Alenia Space using the ESATAN/ESARAD \cite{esatan2003} numerical thermal analysis software. This model has two purposes: (1) to characterize the thermal steady state of each part of the instrument, and (2) to quantify dynamical thermal transfer functions between two physical points of the instrument. Both derive from the need to allow \Planck{} to control thermal systematic effects at the sub-$\mu$K level (see table~\ref{tab:ThermalErrorBudget}). In the specific case of the focal plane stability requirements \cite{2009_LFI_cal_M2}, these call for periodic fluctuations to be less than $\pm 0.9\ \mu$K per pixel in the final maps and for spin-synchronous (s.-s.) variations to be below $\pm 0.45\ \mu$K per pixel. The tighter requirement on s.-s. fluctuations is due to the fact that their period is equal to the spin period of the satellite, i.e. 60\,s, and therefore the associated errors in the maps are not reduced by the redundancy of scanning the same sky circle 60 times. This imposes a high degree of temperature stability at the level of the 20\,K sorption cooler cold end ($\lesssim 100$~mK peak-to-peak).

Due to the criticality in measuring the focal plane thermal stability, 12 silicon diode thermometers are placed on the LFI focal plane (see fig.~\ref{fig:temperatureSensors}). All sensors have dedicated calibration curves and each exhibits an accuracy of $\sim$ 20\ mK. Five of them have been calibrated in the range 14-26.5\ K and the associated readout electronics allow a sensitivity of 0.9\ mK. The others have been calibrated over the 15-90\ K range, with two main sensitivity ranges: 13.3\ mK from 25 to 90\ K and 1.4\ mK from 15 to 25\ K. The intrinsic noise of each thermometer is always below the calibration uncertainties. The characteristics of the thermometers are not sufficient to detect the very slight temperature changes expected during flight, which are estimated by the thermal model to be of the order of 1\,mK peak-to-peak. However, they have been used extensively in this work to measure the response to induced temperature fluctuations on the FPU (much greater than 1\,mK) and will be used during flight to monitor any unexpected change in the temperature of the focal plane.

\section{Estimation of the Focal Plane Transfer Functions and Dynamic Validation of the Numerical Model}
\label{sec:dynamicValidation}

\begin{figure}[tbf]
  \centering
  \includegraphics[width=0.6\textwidth]{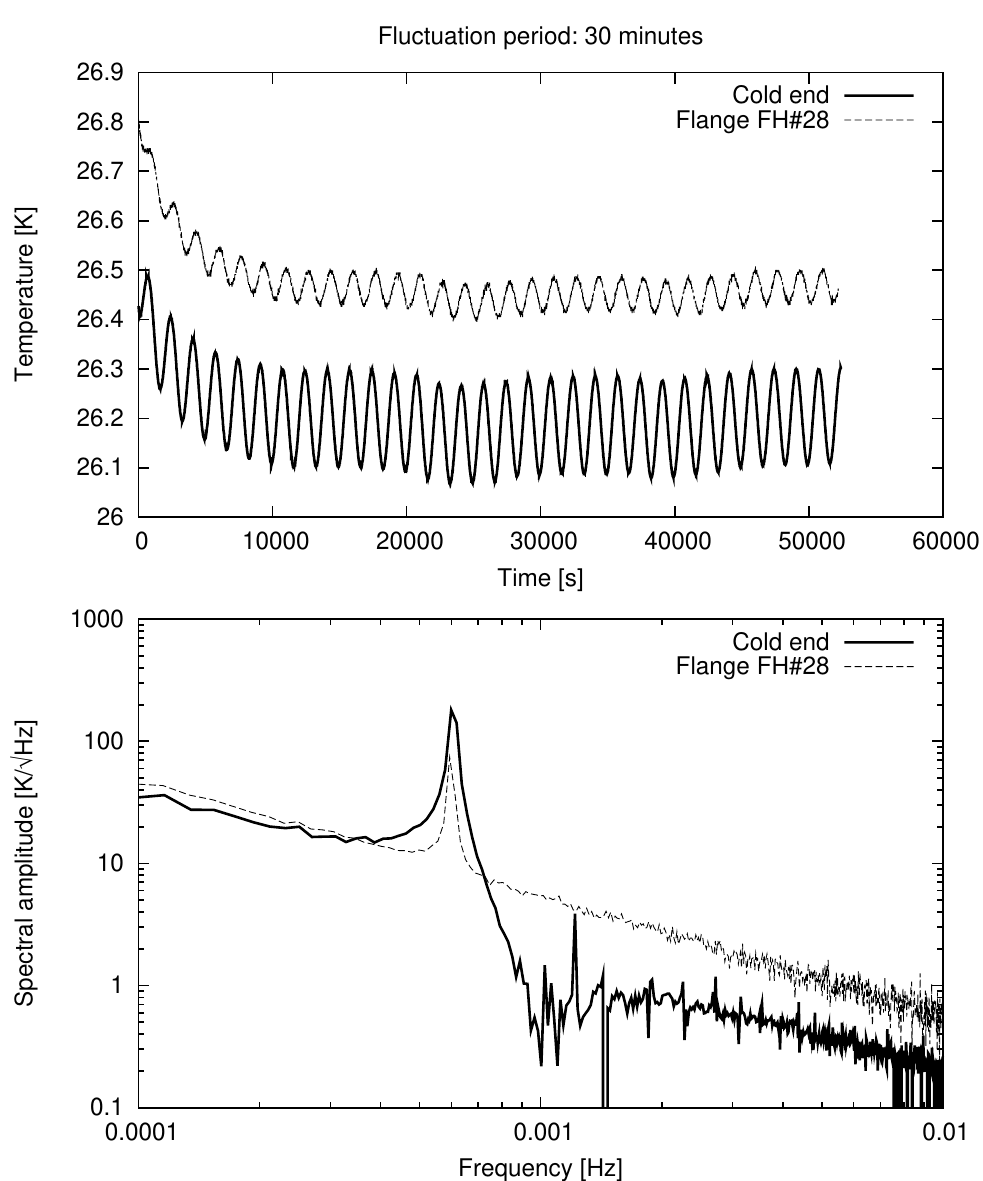}
  \caption{Example of a fluctuation induced at the cold end of the focal plane and the measured temperature response at one of the thermometers (TS5R, placed on the flange of feed horn 28, see Figure~\protect\ref{fig:temperatureSensors}) during the LFI RAA tests. \textbf{Above:} Temperature profiles at the cold end and at the flange. The amplitude of the fluctuation at the flange is smaller than the one at the cold end because of the thermal path. \textbf{Below:} Spectral profile of the two temperature profiles. Each profile has a peak centered at $\nu = 0.55\,\unit{mHz}$, corresponding to a period of 30 minutes. (The $1/f$ component is due to the cold end and not to the intrinsic noise of the thermometers.)}
  \label{fig:thf0004Sample}
\end{figure}

After the integration of the LFI flight instrument, we verified the ESATAN thermal model of the focal plane by forcing thermal fluctuations at the sorption cooler cold end in cryogenic conditions, measuring the induced fluctuations on the thermometers placed on the focal plane (see fig.~\ref{fig:thf0004Sample}) and then comparing the results with the estimates of the thermal model. Such analysis has allowed us to verify the compliancy of the LFI focal plane temperature stability with scientific requirements, as well as to experimentally sample the thermal transfer function for a set of frequencies.

\subsection{Test Methodology}

To perform the measurements, we induced a sinusoidal temperature change of fixed frequency $\nu$, amplitude $\Delta T_0$ and phase $\varphi_0$ (the \emph{input fluctuation}) at the cold end and measured the response at each of the focal plane thermometers.

In principle, any temperature profile can be induced and thermal transfer functions can be derived through a Fourier transform, which allows multiple frequencies to be studied at the same time. However, preliminary studies during the calibration of the LFI Qualification Model (QM) in 2005 proved that variations with a strongly peaked spectrum are easier to analyze and produce more accurate results.

From the time stream of temperatures (one for each thermometer) we calculate the amplitude $\Delta T$ and phase $\varphi$ of the sinusoids and compared them with the input fluctuation. We estimate the $\gamma$ factor as the ratio of the two amplitudes $\Delta T / \Delta T_0$ and the phase shift $\Delta \varphi$ as the difference of the two phases $\varphi - \varphi_0$. By repeating this test with different frequencies $\nu$ for the input fluctuation, we reconstruct the profile of $\gamma (\nu)$ and $\Delta \varphi (\nu)$ as a function of the frequency. The complex quantity $\gamma(\nu) \exp\bigl(i\varphi (\nu)\bigr)$ is called the \emph{thermal transfer function} between the cold end and the thermometer\footnote{It is easy to show that if $\gamma e^{i \varphi}$ is the transfer function from point 1 to point 2, then the t.f.\ from point 2 to point 1 is $\gamma^{-1} e^{-i \varphi}$. We shall use this relation in Sec.~\ref{sec:contactR}.}

During the LFI FM cryogenic tests in the Thales Alenia Space laboratories we performed three separate injections of sinusoidal fluctuations with frequencies 0.18\,mHz, 0.55\,mHz and 1.40\,mHz (corresponding to periods of 12, 30 and 60 minutes) in the cold end. This was done by inducing a time-varying power of sinusoidal shape through a resistor fixed near the cold end.

To derive the transfer function from the thermal model, we apply a sinusoidal temperature profile as a boundary condition and calculated the predicted damping and phase shift.

\subsection{Numerical Methods used in the Analysis}
\label{sec:numericalMethods}

\begin{figure}[tf]
  \centering
  \includegraphics[width=0.75\textwidth]{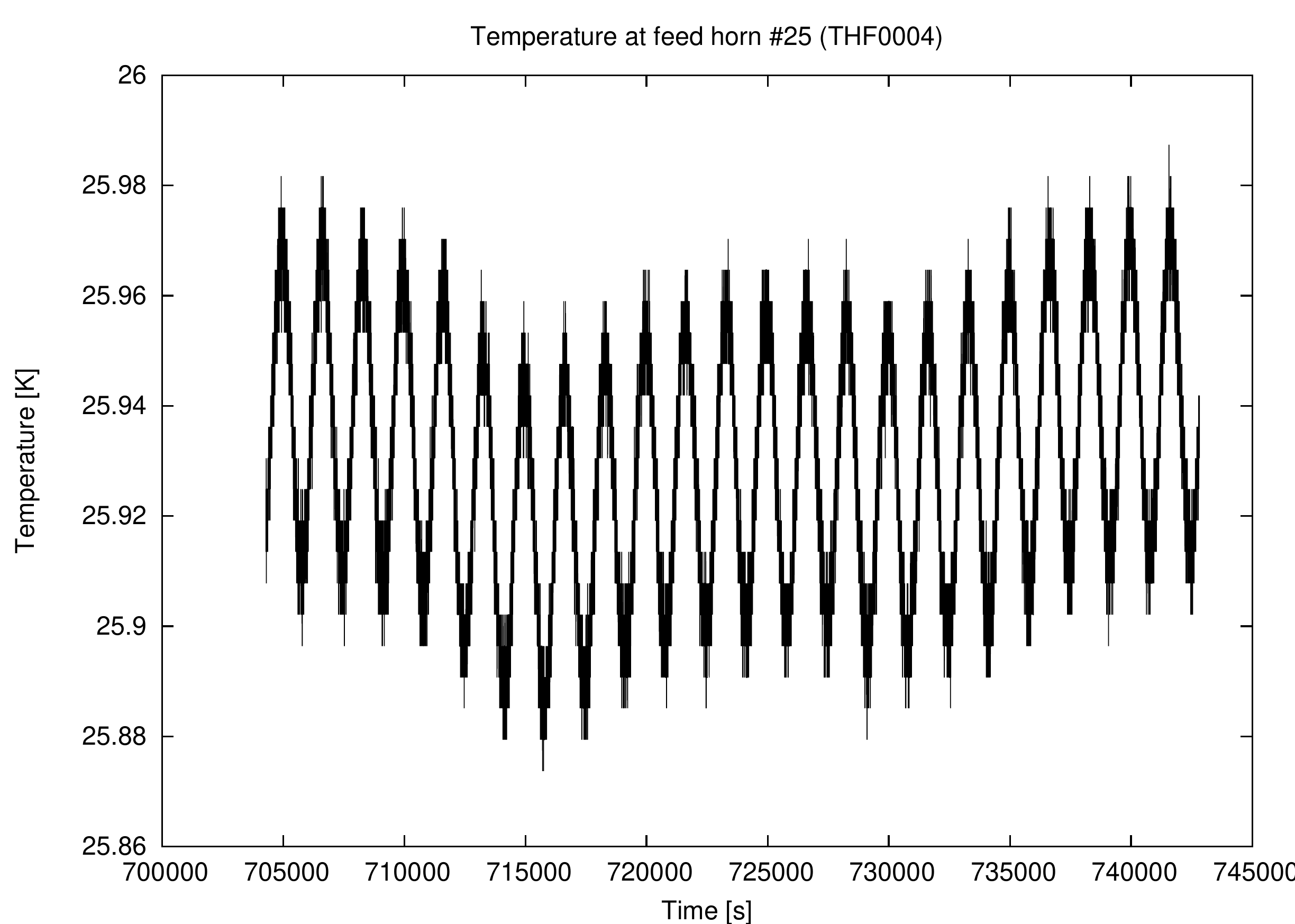}
  \caption{Temperature measured by the thermometer near the LFI feed horn \#25. The temperature sinusoid applied at the cold end had a frequency of 0.55\,mHz. Note the quantization induced by the digital thermometer and the slow thermal drift (longer than one period of the sinusoid). Any method used to extract the amplitude and phase of the sinusoid must be able to limit the impact of both effects on the calculations.}
  \label{fig:thfFluctuation}
\end{figure}

The standard method used to estimate transfer functions is based on a straightforward application of the Fourier Transform. This kind of analysis has the advantage of being quite simple to implement and produces solid results but, on the other hand, it does not provide an easy way to quantify errors in the estimation of the transfer functions. Such errors are generated mainly by the slow temperature drifts, the signal quantization induced by the digital acquisition board (see fig.~\ref{fig:thfFluctuation}) and the intrinsic noise of the thermometers. For this reason, we have developed three different analytical methods that can both reduce the impact of such effects on the final result and quantify the errors of the estimation: (1) a Fourier method which uses a jackknife-like test to derive the error on the estimates of $\gamma$ and $\Delta\varphi$, (2) a method which directly extracts $\gamma$ and $\Delta\varphi$ from time-domain data and (3) a method using non-linear fitting algorithms. The three methods are explained in Appendix~\ref{sec:analysisMethods}. 

To derive our best estimate for $\gamma$ and $\Delta\varphi$ we simply picked the estimate with the lowest relative error (e.g.\ $\delta\gamma/\gamma$) among the one produced by the three methods. In general, the fitting method has produced the best results for $\gamma$ while the Fourier method has outperformed the others in the determination of $\Delta\varphi$.

Calculating $\gamma$ and $\Delta\varphi$ from the numerical thermal model is straightforward, because these temperature streams suffer neither quantization errors nor drifts. Therefore, for the sake of simplicity we have analyzed numerical data only using the time-domain method.

\subsection{Discussion of the Results}
\label{sec:experimentalResultsDiscussion}

The purpose of the validation tests on the LFI focal plane model is to verify that the measured damping level of thermal fluctuations did not overcome the estimate of the numerical model, since the latter has been verified to be compatible with the scientific requirements of LFI.

Of the 12 thermometers in the focal plane, we have chosen to exclude from our analysis TS2L and TS5L because of a few problems occurred with them during the acquisition.

The results for the analysis of the 10 sensors for each of the three frequencies are reported in tables \ref{tbl:FMResults}, \ref{tbl:FMGamma} and \ref{tbl:FMPhi} at the end of the article. The three methods show an excellent agreement (within a few percent) for their estimates and errors on $\nu$ and $\gamma$. We get larger error bars for $\Delta\varphi$ when using the fitting method (see table \ref{tbl:FMPhi}); however, the three methods produce estimates which are always within $2\sigma$, therefore confirming the validity of each method.

The comparison between these values and the estimates of the numerical model is shown in figures \ref{fig:gamma} and \ref{fig:phi}. In these plots $\gamma$ is always smaller in the experimental line than in the numerical estimates, while the contrary applies to $\Delta\varphi$. Both effects were somewhat expected, because the thermal model cannot take into account unknown contact resistance effects. This topic will be addressed in Sec.~\ref{sec:contactR}.


\section{Impact of Instabilities on LFI Maps}
\label{sec:stabilityEstimation}

In this section we use the measured transfer functions to quantify how the better damping (i.e.\ smaller values for $\gamma$) will reduce the error caused by sorption cooler instabilities in the LFI maps. To do this, we need to simulate the propagation of fluctuations at the sorption cooler cold end to the production of the maps. The stpng involved in this calculations are the following (taken from \cite{2002A&A...384..736M}):
\begin{enumerate}
\item the temperature fluctuation propagates from the cold end to the radiometer in a way described by the thermal transfer functions discussed in this paper;
\item the fluctuation in the radiometer temperature induces a fluctuation in its output which is proportional to the temperature itself \cite{2009_LFI_cal_R6};
\item in the production of the map, multiple passes over the same pixel are averaged, thus further reducing the error caused by non-spin synchronous fluctuations;
\item destriping techniques are going to be applied to the Planck maps in order to reduce the impact of long-term drifts, which include sorption cooler fluctuations as well.
\end{enumerate}

\subsection{Assumptions Used in the Analysis}

\begin{table}
	\centering
	\begin{tabular}{|c|c|l|}
	\hline
	Parameter & Value & Notes \\
	\hline
	$L_\text{fh-OMT}$            & 0.25\,dB  & Measured at room temperature \\
	$L_\text{4K}$                & 0.25\,dB  & Measured at room temperature \\
	$T_\text{sky}$               & 3.7\,K    & 2.7\,K from the sky plus 1\,K from the telescope \\
	$T_\text{4K}$                & 4.5\,K    & Estimated value \\
	$T_\text{phys}^\text{FEM}$ & 26\,K     & From table 3 in \cite{2009_LFI_cal_M3} \\
	$T_n$                          & 10.5\,K   & From table 1 in \cite{2009_LFI_cal_R2} (mean between 28-M0 and M1) \\
	$\frac{\partial G}{\partial T_\text{phys}^\text{FEM}}$
	                               & -0.03\,dB & From table 11 in \cite{2009_LFI_cal_R6} \\
	$\frac{\partial T_n}{\partial T_\text{phys}^\text{FEM}}$
	                               & 0.15\,dB  & From table 11 in \cite{2009_LFI_cal_R6} \\
	$r$                            & 0.95      & Estimated value \\
	\hline
	$T_f^\text{FEM}$             & 1.05$\times 10^{-2}$ & From eq.~\ref{eq:radiometricTF} \\
	\hline
	\end{tabular}
	\caption{\label{ref:radiometricTFParams} List of parameters used in eq.~\protect\ref{eq:radiometricTF} to estimate the radiometric transfer function for radiometer \#28-M (30\,GHz).}
\end{table}

We have studied the case for radiometer \#28-M (30\,GHz), as horn \#28 is the easiest to use for our study because of the temperature sensor mounted on its flange (sensor TS5R, fig.~\ref{fig:temperatureSensors}). To carry out the calculations, we use the following assumptions:
\begin{itemize}
\item The radiometric response $T_f^\text{FEM}$ is defined such that the temperature change on the map $\delta T_\text{map}$ due to a variation of the physical temperature $\delta T_\text{phys}$ is $\delta T_\text{map} = T_f^\text{FEM} \times \delta T_\text{phys}$. We use eq.~2.5 in \cite{2009_LFI_cal_R6} to estimate $T_f^\text{FEM}$ analitically, with the simplifying assumptions of a perfect match in the two gains ($G := G_{F1} = G_{F2}$) and noise temperatures ($T_n := T_{nF1} = T_{nF2}$):
\begin{equation}
\label{eq:radiometricTF}
\begin{split}
T_f^\text{FEM} = L_\text{fh-OMT} \times \Biggl( & \left(1 - \frac{1}{L_\text{fh-OMT}}\right) - r \left( 1 - \frac{1}{L_\text{4K}}\right) + \\
& \left[\tilde T_\text{sky} + T_n - r (\tilde T_\text{4K} + T_n)\right] \frac{\partial G}{\partial T_\text{phys}^\text{FEM}} + (1 - r) \frac{\partial T_n}{T_\text{phys}^\text{FEM}}\Biggr),
\end{split}
\end{equation}
with
\begin{eqnarray*}
\tilde T_\text{sky} &=& \frac{T_\text{sky}}{L_\text{fh-OMT}} + \left(1 - \frac{1}{L_\text{fh-OMT}}\right) T_\text{phys}^\text{FEM}, \\
\tilde T_\text{4K} &=& \frac{T_\text{4K}}{L_\text{4K}} + \left(1 - \frac{1}{L_\text{4K}}\right) T_\text{phys}^\text{FEM}. \\
\end{eqnarray*}
The list of parameters used in the estimation of the radiometric t.f.\ are listed in table~\ref{ref:radiometricTFParams}.

\item The measurement redundancy is estimated using an analytical model instead of considering the full details of the Planck scanning strategy. This model only estimates the redundancy of the pixels on the ecliptic plane, which is the worst case. The transfer function is the following (taken from \cite{2002A&A...384..736M}, eq.~3):
\begin{equation}
T_f^\text{map} (\nu) = \frac{2}{N} \left| \frac{\sin(\pi N \nu / \nu_\text{spin})}{\sin (\pi \nu / \nu_\text{spin})} \right|,
\end{equation}
where $N$ is the number of scan circles over which the average of each pixel is computed (in this context we use $N = 24\times 60 = 1440$, corresponding to a 24-hour scanning period over the same circle) and $\nu_\text{spin} = 1/60\,\text{Hz}$ is the Planck spin frequency.

\item The impact of destriping on the maps is estimated via a transfer function, instead of applying the full algorithm to the map data. We use the following transfer function (taken from \cite{2002A&A...384..736M}, fig.~5):
\begin{equation}
T_f^\text{destr} (\nu) = \frac{a}{\nu} + b,
\end{equation}
with $a = 0.1067\,\text{Hz}$ and $b = 1.7992$.

\end{itemize}

With such assumptions, a fluctuation amplitude $\delta T_\text{cold-end}$ with frequency $\nu$ at the cold end induces an error in the map equal to
\begin{equation}
\label{eq:overallTF}
\delta T_\text{map} = \left( T_f^\text{destr} (\nu) \ T_f^\text{map} (\nu) \ T_f^\text{FEM} \ \gamma(\nu) \right) \times \delta T_\text{cold-end},
\end{equation}
with $\gamma(\nu)$ being the thermal transfer function between the cold end and the radiometer. As said above, in our case we shall use the values of $\gamma$ reported in table~\ref{tbl:FMResults} for sensor TS5R.

\subsection{Study of Sinusoidal Fluctuations}

In this paragraph we use the algorithms and formulae described above to study how the fluctuation amplitudes reported in table~\ref{tab:ErrorBudget} change when switching from the numerical to the measured transfer functions. Our approach follows these stpng:
\begin{enumerate}
\item By inverting the whole process, we derive a fluctuation amplitude at the cold end that produces an error in the map equal to the number reported in table~\ref{tab:ErrorBudget} when the \emph{numerical} t.f.\ is used;
\item We study the error on the map caused by the same fluctuation when the \emph{measured} t.f.\ is applied.
\end{enumerate}
We have chosen not to do this comparison for spin-synchronous and high-frequency fluctuations, as our measurements do not cover such frequency ranges. Instead we concentrated on a range of frequencies similar to the ones measured during the instrument tests.

\begin{table}
	\centering
	\begin{tabular}{|ccccc|}
	\hline
	$\nu$ & $\delta T_\text{cold end}$ & $\delta T_\text{map}^\text{num}$ & $\delta T_\text{map}^\text{meas}$ & Improvement \\
	\hline
	0.25\,mHz & 0.339\,mK & 0.900\,$\mu$K & 0.526\,$\mu$K & 41\% \\
	0.60\,mHz & 0.244\,mK & 0.900\,$\mu$K & 0.549\,$\mu$K & 39\% \\
	1.50\,mHz & 0.257\,mK & 0.900\,$\mu$K & 0.539\,$\mu$K & 40\% \\
	\hline
	\end{tabular}
	\caption{\label{tab:fluctFromRequirements} Difference between the peak-to-peak fluctuation in the LFI temperature maps estimated using the numerical ($\delta T_\text{map}^\text{num}$) and the measured ($\delta T_\text{map}^\text{meas}$) thermal t.f.\ for three frequencies. The 0.25\,mHz and 1.5\,mHz frequencies are respectively the average frequency of each of the six sorption cooler compressor elements cycle and of the overall compressor cycle, and they are therefore of considerable importance in the analysis of LFI systematics. The fluctuation amplitude at the cold end $\delta T_\text{cold end}$ has been chosen so that a 24-hour observation leads to $\delta T_\text{map}^\text{num} = 0.900\,\mu\text{K}$. The calculation has been done for radiometer \#28-M (30\,GHz).}
\end{table}

Table~\ref{tab:fluctFromRequirements} reports the results of our calculations: the reduction in the value of $\gamma (\nu)$ from the numerical to the measured numbers (between 39\% and 44\%, see fig.~\ref{fig:gamma}) has lead to a comparable improvement in the magnitude of the error in the map, between 39\% and 41\%.






\section{Estimation of the Impact of Contact Resistances in the Measurements}
\label{sec:contactR}

The discrepancies between the model and the experimental measurements in general can be explained by the presence of non-ideal phenomena that are often very hard, if not impossible, to simulate. Among these phenomena, in a thermal system, the contact resistance is usually one of the most important. In this section we shall expand our idea proposed in Sec.~\ref{sec:experimentalResultsDiscussion} that the differences between the model and the data are due to the presence of unknown contact resistances in the system and in particular at the main interface between the sorption cooler cold end and the LFI focal plane.

A contact resistance is a constant that characterizes a non-ideal thermal contact between two bodies. It is expressed as the ratio between the temperature jump $\Delta T$ at the interface and the flux per unit area $q$ across the interface itself:
\begin{equation}
\label{eq:rc}
R_c = \frac{\Delta T}{q}.
\end{equation}
As shown in Appendix~\ref{sec:contactResistances}, when a sinusoidal fluctuation propagates through the interface (the case of interest in the context of this paper) there is a sharp decrease in the value of the fluctuation amplitude $\gamma$ and an increase in the value of the phase shift $\Delta \varphi$ (see fig.~\ref{fig:contactResistance}). Note that this is in agreement with our comparison between estimated and measured transfer functions, as the measurements always showed greater damping (smaller $\gamma$, see fig.~\ref{fig:gamma}) and larger phase shift (larger $\Delta\varphi$, see fig.~\ref{fig:phi}).

The experimental measurement of contact resistances is a complex task which requires dedicated instrumental set-ups and is therefore not easily applicable in the context of the Planck/LFI focal plane thermal tests, where a limited number of temperature sensors are available and the involved objects have complex shapes. From eq.~\ref{eq:rc}, in order to estimate $R_c$ the temperature change at the interface and the heat flux flowing through the interface must be estimated. Both quantities are not measurable directly but must be extrapolated from a number of temperatures measurements near the interface\footnote{An additional complication is that the mathematical problem is ill-posed and requires the use of inverse method, which often suffer from numerical instabilities. See \cite{2000OzisikInverseHeatTransfer}.}. Although it is sometimes possible to estimate $R_c$ from transient temperature measurements (see e.g.\ \cite{2008FiebergContactResistance}), the most common set-up uses bodies in thermal equilibrium in the one-dimensional approximation, since this configuration is easily modeled mathematically (see Appendix~\ref{sec:crStaticCase}). In any case, for the mathematical problem to be solvable it is required to measure the temperature at a number of points near both sides of the interface: a quick glance at fig.~\ref{fig:temperatureSensors} reveals that this is very hard to accomplish with the Planck/LFI focal plane thermometers.

To increase our confidence in explaining the discrepancies observed in the measured vs.\ estimated transfer functions, we have therefore chosen to perform a statistical analysis of such discrepancies. We have estimated all the possible transfer functions between two sensors of the focal plane or the sensor at the cold end (both numerically and from the same experimental data discussed in the previous sections) and have divided them into two groups:
\begin{description}
\item[Focal plane t.f.:] These are the transfer functions between two thermometers on the focal plane, i.e.\ pairs of the thermometers shown in fig.~\ref{fig:temperatureSensors}. Since there are 10 thermometers available, this leads to $10 \times 9 = 90$ transfer functions. For the sake of simplicity we have estimated the t.f.\ using the time-domain method (see Appendix~\ref{sec:analysisMethods}) instead of applying the three methods and picking the best estimate.

\item[Cold end t.f.:] These are the transfer functions between the cold end sensor and one thermometer on the focal plane. They have been derived and discussed in Sec.~\ref{sec:dynamicValidation}.
\end{description}
Since non-ideal thermal contacts are not considered by the numerical model, if we expect that an important non-ideal contact be at the cold end/focal plane interface then the discrepances between the numerical and the measured t.f.\ should be \emph{larger} for the cold end transfer functions than for the focal plane t.f.  We quantify the discrepance between two different estimates of a t.f.\ (i.e.\ numerical vs.\ measured) in terms of two quantities:
\begin{eqnarray}
\delta_\gamma (\nu) &=& 2 \frac{\left|\gamma_\text{meas}(\nu) - \gamma_\text{num}(\nu)\right|}{\gamma_\text{meas}(\nu) + \gamma_\text{num}(\nu)}, \label{eq:deltaGamma} \\
\delta_\varphi (\nu) &=& \frac{\left|\Delta\varphi_\text{meas}(\nu) - \Delta\varphi_\text{num}(\nu)\right|}{2\pi}, \label{eq:deltaPhi}
\end{eqnarray}
where $\bigl(\gamma_\text{meas}(\nu), \Delta\varphi_\text{meas}(\nu)\bigr)$ is the measured t.f.\ and $\bigl(\Delta\varphi_\text{num}(\nu), \Delta\varphi_\text{num}(\nu)\bigr)$ is the numerical estimate. (With this definition of $\delta_\gamma$ and $\delta_\varphi$, the two quantities do not change when one exchanges the two sensors, i.e. $\delta_\gamma^{1\rightarrow 2} = \delta_\gamma^{2 \rightarrow 1}$ and $\delta_\varphi^{1\rightarrow 2} = \delta_\varphi^{2 \rightarrow 1}$.) In principle each quantity is able to estimate the discrepancy between two t.f.  However, we expect the presence of the contact resistance to have a greater effect in $\delta_\gamma$ than in $\delta_\varphi$ (see Sec.~\ref{sec:crDynamicCase} in the appendix and expecially fig.~\ref{fig:contactResistance}: compare e.g.\ the jump in $\gamma(\nu)$ and in $\Delta\varphi(\nu)$ for $R_c = 1$).

\begin{figure}[tbf]
	\centering
	\includegraphics[height=0.9\textheight]{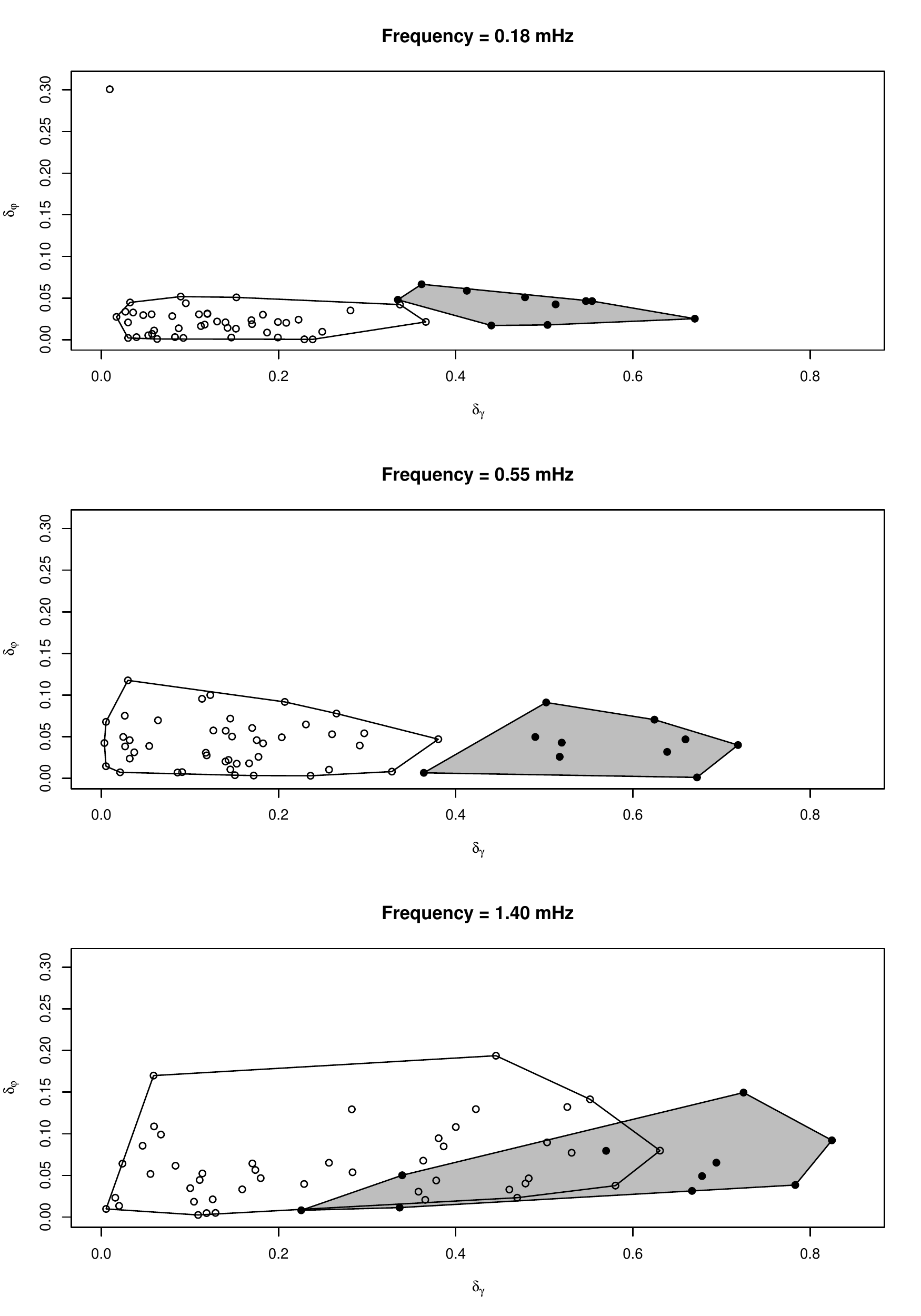}
	\caption{\label{fig:bivariateGammaPhi} Bivariate scatter plot of the quantities $\delta_\gamma$ and $\delta_\varphi$ (equations~\protect\ref{eq:deltaGamma} and \protect\ref{eq:deltaPhi}) for 10 out of 12 focal plane thermometers. They have been grouped according to the three frequency measured during the instrument tests. White points have been calculated using the transfer functions between sensors on the focal plane ($10 \times 9 / 2 = 45$ points), while black points have been calculated using the t.f.\ between the cold end sensor and each of the 10 focal plane thermometers used in this work. The white and grey polygons are the convex hulls of the two point sets. The discussion of these plots is provided in the text.}
\end{figure}

\afterpage{\clearpage}

The result of our analysis is shown in fig.~\ref{fig:bivariateGammaPhi}, which shows the points with coordinates $(\delta_\gamma, \delta_\varphi)$ for each of the three frequencies and each of the focal plane t.f.\ (white circles) and the cold end t.f.\ (black circles). The convex hulls\footnote{The \emph{convex hull} of a set of points is the convex polygon with the smallest area that encloses all the points in the set.} of the two sets of points are shown as well. It is evident that if we consider the $\delta_\gamma$ parameter (abscissa), then discrepancies in cold-end t.f.\  are systematically higher than those in focal plane t.f.: this is exactly what we were expecting from our hypotheses, and thus confirms the fact that the differences between the measurements and the numerical estimates in figures~\ref{fig:gamma} and \ref{fig:phi} are likely to be due to a non-ideality at the SCS cold end/LFI focal plane discontinuity.


\section{Conclusions}
\label{sec:conclusions}

We have presented the results of our analysis of the thermal performance of the focal plane of the \Planck{}/LFI instrument. The purpose of this work is twofold: (1) to experimentally measure the thermal transfer function between the 20\,K focal plane cold end and a set of points in the focal plane itself, and (2) to use such measurements to validate the numerical thermal model of the focal plane. The latter point is of capital importance for LFI, since the design of the instrument has been driven by the thermal stability estimates produced by the model itself.

We have estimated the focal plane thermal transfer functions by inducing a sinusoidal temperature variation near the cold end while measuring the impact on the thermal stability on other points of the focal plane. We have applied three different analysis methods which have provided results in good agreement, especially in estimating the amplitude reduction (the $\gamma$ factor).

Our analysis has compared measured transfer functions with simulations produced via the numerical thermal model of the LFI focal plane. We have found that in every case the thermal mass of the instrument is able to damp fluctuations better than predicted by the model, probably because of a non-ideal contact between the focal plane and the cold end. Therefore, this analysis not only confirms the LFI thermal performance compliancy to the expected sensitivity per pixel level of $\Delta T/T \sim 10^{-6}$ in the final maps, but in some cases estimates an improvement of $\sim$40\% in the peak-to-peak effect of long-term focal plane temperature fluctuations on the maps produced by LFI.

The next step of our study is to assess the impact of the residual effect of such fluctuations on the science, as well as to characterize the impact of fluctuations on the LFI focal plane induced by thermal instabilities originating within HFI, the other instrument on board of \Planck{}. Both activities will be possible by exploiting data taken during integrated satellite tests at cryogenic conditions on ground and during the early flight phases, currently being analyzed.


\acknowledgments

\Planck{} is a project of the European Space Agency with instruments funded by ESA member states, and with special contributions from Denmark and NASA (USA). The \Planck{}/LFI project is developed by an International Consortium lead by Italy and involving Canada, Finland, Germany, Norway, Spain, Switzerland, UK, USA. The Italian contribution to \Planck{} is supported by the Italian Space Agency (ASI).

This work has been done under an ASI contract \Planck{}/LFI Activity of Phase E2.

The authors would like to express their gratitude to Dr.\ Stuart Lowe for his help in correcting the text.

\bibliographystyle{plainnat}
\bibliography{thermal-model,references_prelaunch_forJI}

\appendix

\section{Analysis Methods}
\label{sec:analysisMethods}

Here we provide a short description of the three analysis methods used to extract the cold-end thermal transfer functions to the focal plane thermometers.

\begin{figure}[tbf]
  \centering
  \includegraphics[width=0.75\textwidth]{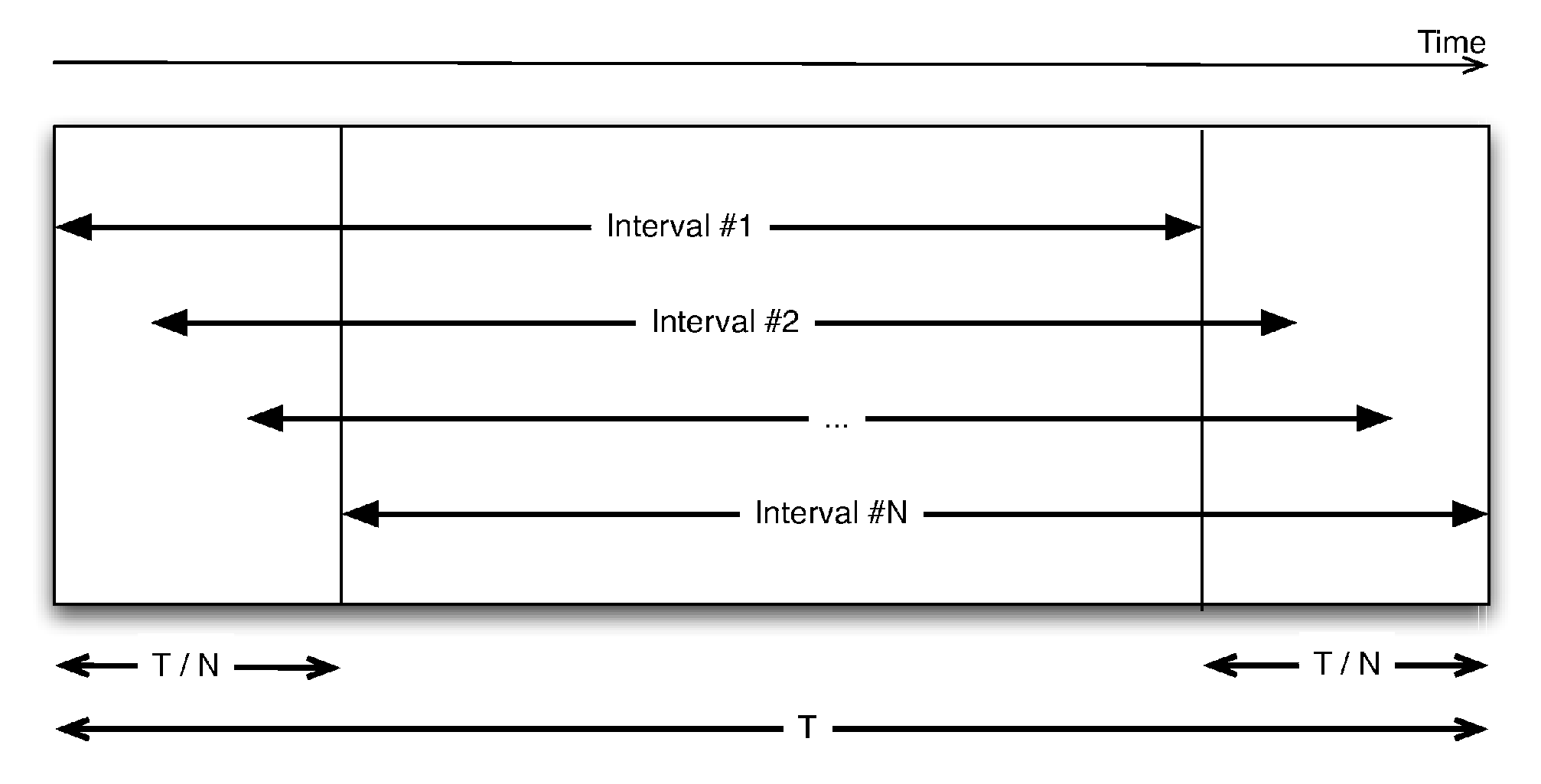}
  \caption{\label{fig:timeWindows} Layout of the time windows used in the Fourier method. Within a time window of length $T$, we consider $N$ intervals of length $(N-1)/N\times T$. Each interval $j=1\ldots N$ starts at $(j-1)/(N-1)\times T/N$.}
\end{figure}

\subsection{Fourier method}

The Discrete Fourier Transform (DFT) of the temperature streams at the two points is calculated after the two datasets have been interpolated to the same sampling frequency. The value of $\gamma$ is given by the ratio in the height of the peaks in the two transforms, while the estimate of the phase delay $\Delta\varphi$ is given by the difference in the phases of the two peaks.

As said above, getting a meaningful estimate of the error of $\gamma$ and $\Delta\varphi$ from this method is not trivial. We calculated $\nu$, $\gamma$ and $\Delta\varphi$ over $N$ partially overlapping time windows, thus obtaining $N$ estimates for each quantity. We then considered our best estimate and its error as the mean value and the standard deviation of the set, respectively\footnote{We chose not to divide this error by $\sqrt{N}$ because the $N$ estimates are not truly independent, as they are calculated over time windows which partially overlap.}. A sketch showing how the $N$ time windows overlap is illustrated in fig.~\ref{fig:timeWindows}.

\subsection{Time-domain method}

After having downsampled\footnote{The frequency of the downsampled data is chosen iteratively, with the goal of optimizing the relative error on the estimates for $\gamma$ and $\Delta\varphi$.} the data, the peaks of the sinusoidal temperature streams at the two points are found numerically. From each consecutive pair of peaks an estimate of frequency $\nu$, damping $\gamma$ and phase shift $\Delta\varphi$ can be easily derived. Assuming that we have $N$ pairs, the best values for $\nu$, $\gamma$ and $\Delta\varphi$ are simply the average of the $N$ estimates. Basic error propagation is used to derive estimates for the errors associated with each quantity.

This method has the advantage of providing a straightforward way to predict the order of magnitude of the expected error on the measurement of $\gamma$ and $\Delta\varphi$ before actually doing the calculations. As an example, we can derive the error on the $\gamma$ parameter considering the accuracy of the focal plane thermometers: if this accuracy is 20\,mK, then the fluctuation amplitude (difference between two temperatures) has an error of
\[
\frac{\sqrt{2 \times (20\,\text{mK})^2}}{\sqrt{2N}} = \frac{20\,\text{mK}}{\sqrt{N}},
\]
where $N$ is the number of periods and is typically between 15 and 30. Since the absolute level of the temperature during the RAA tests was between 20 and 30\,K, this means that the relative error on the amplitude is of the order of $10^{-3}$: this is roughly the same as the expected error for $\gamma$, since this quantity is a ratio between two amplitudes. These are indeed the errors we get on this quantity from the analysis of the \Planck/LFI RAA tests (see table~\ref{tbl:FMGamma}).

\subsection{Fitting method}

Temperature profiles are fitted numerically with a function of the form $T \sin (2\pi\nu t + \theta) + p(t)$, where $p(t)$ is a polynomial which models slow temperature drifts. From this fit the values for $\gamma$ and $\Delta\varphi$ as well as their errors can be easily estimated.

\section{Analytical Models of Contact Resistances}
\label{sec:contactResistances}

In this appendix we discuss two simple analytical models that show the effects due to a non-ideal thermal contact between two bodies (hereafter named ``body 1'' and ``body 2''). We only use dimensionless quantities for the sake of simplicity. The bodies are supposed to be monodimensional along the $x$ direction and connected at $x = 1$, and their thermal properties (heat capacity, density and thermal conductivity) do not depend on $x$ nor on time $t$. The governing equations for the temperatures $T_1$ and $T_2$ in the two bodies are:
\begin{eqnarray}
\partial^2_{xx} T_1 (x, t) &=& \partial_t T_1 (x, t), \label{eq:crI}\\
d\,\partial^2_{xx} T_2 (x, t) &=& \,\partial_t T_2 (x, t), \label{eq:crII} \\
-R_c\,\partial_x T_1 (1, t) &=& T_1 (1, t) - T_2 (1, t), \label{eq:crIII} \\
-k\,R_c\,\partial_x T_2 (1, t) &=& T_1 (1, t) - T_2 (1, t), \label{eq:crIV}
\end{eqnarray}
where $d$ and $k$ are the thermal diffusivity and conductivity of the second body with respect to the first and $R_c$ is the contact resistance. Equation \ref{eq:crI} and \ref{eq:crII} are the heat equation expressed for each of the two bodies, equations \ref{eq:crIII} and \ref{eq:crIV} use the definition of contact resistance, which is the ratio per unit surface of the temperature jump at the boundary over the heat flux $q = -k \partial_x T$.

\subsection{Static case}
\label{sec:crStaticCase}

\begin{figure}[tbf]
	\centering
	\includegraphics{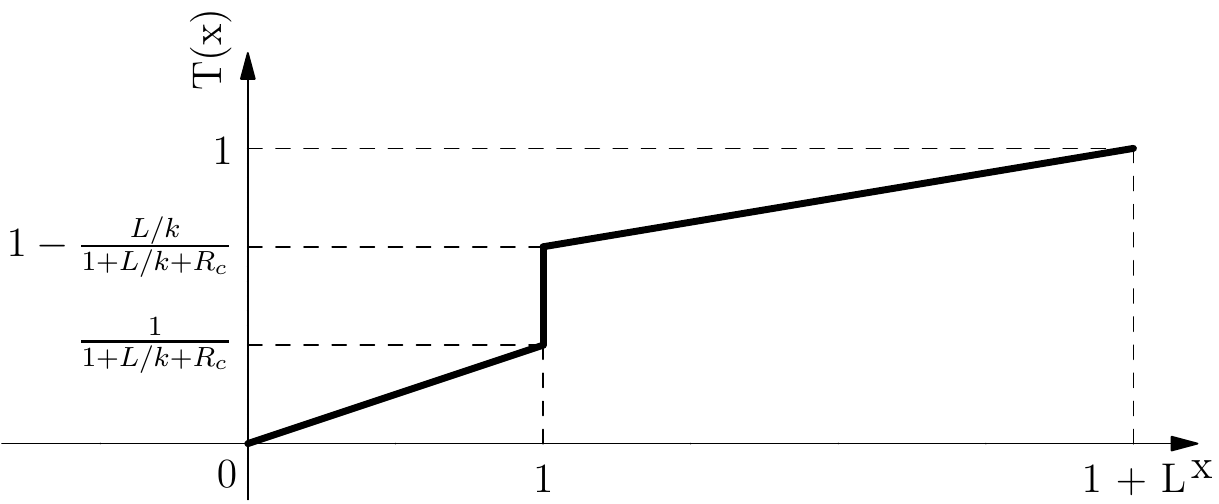}
	\caption{\label{fig:staticResistance} Temperature profile for a two-body system in equilibrium extending from $x = 0$ to $x = L$ and with a contact resistance $R_c$ at $x = 1$ (see eq.~\protect\ref{eq:staticI} and \protect\ref{eq:staticII}).}
\end{figure}

A very simple solution for the set of equations \ref{eq:crI}-\ref{eq:crIV} is found in the steady-state case ($\partial_t T_1 = \partial_t T_2 = 0$). If body 1 extends from $x = 0$ to $x = 1$, and body 2 has a length $L$, then from the boundary conditions $\left. T_1 \right|_{x = 0} = 0$ and $\left. T_2 \right|_{x = 1 + L} = 1$ we derive the following solutions:
\begin{eqnarray}
T_1 (x) &=& \frac{k}{k (1 + R_c) + L} x, \label{eq:staticI}\\
T_2 (x) &=& 1 - \frac{L + 1 - x}{k(1 + R_c) + L}, \label{eq:staticII}
\end{eqnarray}
where $T_1 (x)$ is defined for $x \in [0, 1)$ and $T_2 (x)$ for $x \in (1, 1 + L]$, i.e.\ there is a sudden jump in the temperature across the boundary, and the temperature varies linearly in the two bodies with a slope which depends  on the thermal conductivity of the body. This solution is plotted in fig.~\ref{fig:staticResistance}.

\subsection{Dynamic case}
\label{sec:crDynamicCase}

In analogy with the Planck/LFI tests discussed in this paper, we consider here a sinusoidal temperature fluctuation propagating through the two bodies. The fluctuation is applied to body 1 at $x = 0$, and we assume that body 2 is infinitely long\footnote{The boundary condition for $T_1$ has a minus sign in the exponential in analogy with the standard way to write plane waves, $\exp i(k x - \omega t)$.}:
\begin{eqnarray}
T_1 (x = 0, t) &=& T_a e^{-2\pi i \nu t}, \label{eq:boundaryI} \\
\lim_{x \rightarrow +\infty} \left|T_2 (x, t)\right| &=& 0, \label{eq:boundaryII} 
\end{eqnarray}
with $T_a$ the temperature fluctuation amplitude.

\begin{figure}[tbf]
	\centering
	\includegraphics{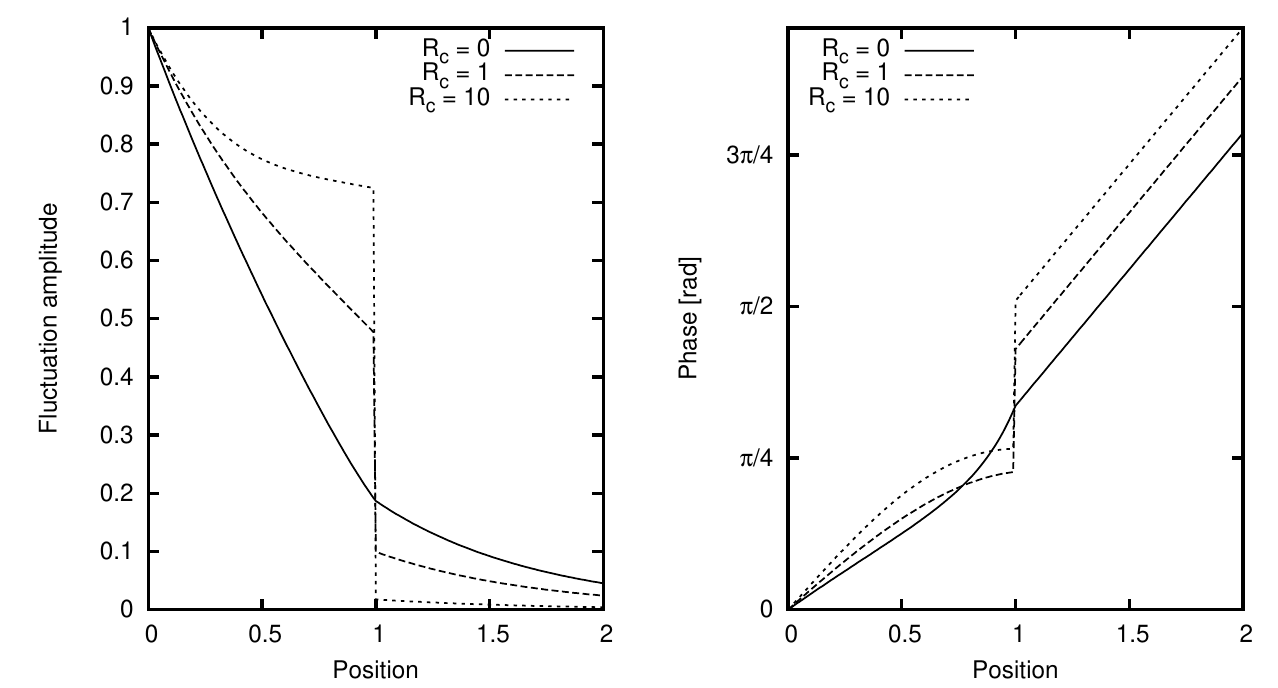}
	\caption{\label{fig:contactResistance} Fluctuation amplitude ($\left|g_{1/2} (x)\right|$) and phase ($\arg g_{1/2} (x)$) for a sinusoidal fluctuation propagating from $x = 0$ and encountering a discontinuity with contact resistance $R_c$ at $x = 1$ (equations \protect\ref{eq:resistSolI} and \protect\ref{eq:resistSolII}). Three values of $R_c$ have been used to produce the plots. The values used for the parameters are $\alpha = 1$, $k = 2$, $d = 1/2$.}
\end{figure}

We write the generic solution $T_{1/2}$ as
\[
T_{1/2} (x, t) = T_a\,g_{1/2} (x)\,e^{-2\pi i \nu t}.
\]
Under this hypothesis, $\left|T_a\, g_{1/2} (x)\right|$ is the fluctuation amplitude of the temperature at point $x$ and $\arg g_{1/2} (x)$ is the phase of the fluctuation. Equations from \ref{eq:crI} to \ref{eq:crIV} rewrite as follows:
\begin{eqnarray}
g_1'' (x) &=& 2\pi i \nu g_1 (x), \\
d\,g_2'' (x) &=& 2\pi i \nu g_2 (x), \\
-R_c g_1' (1) &=& g_1 (1) - g_2 (1), \\
-k R_c g_2' (1) &=& g_1 (1) - g_2 (1).
\end{eqnarray}
Putting $\alpha = \sqrt{\pi \nu}$, $\xi = R_c \sqrt{\pi \nu}$, we get the solutions
\begin{eqnarray}
g_1 (x) &=& c_1 \exp\Bigl(\alpha (1 + i) x\Bigr) + (1 - c_1) \exp\Bigl(-\alpha (1 + i) x\Bigr), \label{eq:resistSolI}\\
g_2 (x) &=& c_2 \exp\left(-\frac{\alpha}{\sqrt{d}} (1 + i) x\right), \label{eq:resistSolII}
\end{eqnarray}
where
\begin{eqnarray}
c_1 &=& \frac{(1+i) \xi k - k + \sqrt{d}}{(1+i) \xi k - k + e^{2\alpha (1 + i)} \left((1+i) \xi k + k + \sqrt{d}\right)+\sqrt{d}}, \label{eq:RCconstI} \\
c_2 &=& \frac{2 \sqrt{d} \exp\left(\alpha (1+i) \left(1+\frac{1}{\sqrt{d}}\right) \right)}{(1+i) \xi  k-k+e^{2\alpha (1 + i)} \left((1+i) \xi k+k+\sqrt{d}\right)+\sqrt{d}}. \label{eq:RCconstII}
\end{eqnarray}
The amplitude and phase of the fluctuation in the two bodies as a function of $x$ are shown in fig.~\ref{fig:contactResistance}. At the boundary ($x = 1$) the ratio between $g_1(x)$ and $g_2(x)$ is the complex number
\begin{equation}
\frac{\lim_{x\rightarrow 1^+} g_2(x)}{\lim_{x \rightarrow 1^-} g_1(x)} = \frac{\sqrt{d}}{k \xi}\,\frac{1}{\sqrt{d}/(k \xi) + 1 + i},
\end{equation}
whose absolute value $\gamma_\text{Rc}$ and argument $\Delta\varphi_\text{Rc}$ are
\begin{eqnarray}
\gamma_\text{Rc} &=& \frac{\sqrt{d}}{k \xi}\sqrt{\frac{1}{\left(\sqrt{d}/(k \xi) + 1\right)^2 + 1}}, \label{eq:RCdamping} \\
\Delta\varphi_\text{Rc} &=& -\arctan\left(\frac{1}{\sqrt{d}/(k \xi) + 1}\right). \label{eq:RCshift}
\end{eqnarray}
The value of $\gamma_\text{Rc}$ is the ``amplitude drop'' at $x = 1$ (left side of fig.~\ref{fig:contactResistance}) and the value of $\varphi_\text{Rc}$ is the phase shift at the same point (right side of the same figure). Therefore, fig.~\ref{fig:contactResistance} shows that a non-ideal thermal contact between two bodies has the effect of damping the amplitude of thermal fluctuations and inducing a phase delay in the propagating wave. Both effects increase for large values of $R_c$ (i.e.\ if $R_c\rightarrow +\infty$ then $\gamma_{Rc} \rightarrow 0$ and $\Delta\varphi \rightarrow -\pi/4$).

It is easy from these formulae to obtain the expressions with the proper dimensions. If the two bodies have thermal diffusivity coefficients $D_1$ and $D_2$ and thermal conductivities $k_1$ and $k_2$, and the discontinuity is at $x = L$, it is enough to use the following substitutions in equations from \ref{eq:resistSolI} to \ref{eq:RCshift}:
\[
d = \frac{D_2}{D_1}, \quad k = \frac{k_2}{k_1}, \quad \alpha = \sqrt{\frac{\pi \nu}{D_1}}, \quad \xi = R_c\,A\,\sqrt{\frac{\pi \nu}{D_1}},
\]
where $A$ is the area of the boundary surface at $x = L$. Moreover, in equations \ref{eq:RCconstI} and \ref{eq:RCconstII} the substitution
\begin{equation}
\alpha = L \sqrt{\frac{\pi \nu}{D_1}}
\end{equation}
must be made in the exponentials.

\begin{figure}[f]
  \centering
  \includegraphics[width=0.8\textwidth]{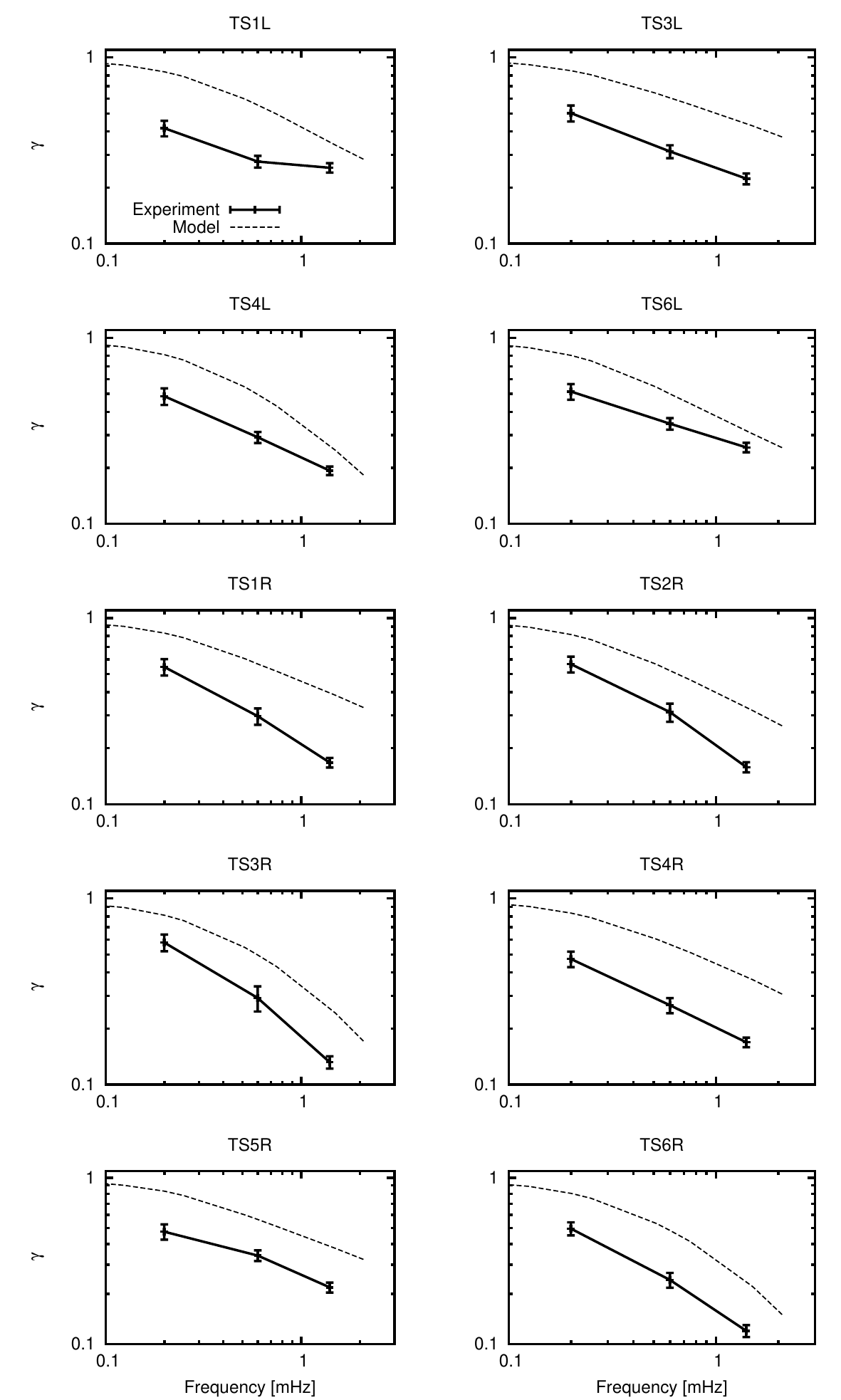}
  \caption{Comparison between the estimates of the numerical model and the experimental results for the $\gamma$ function (damping). X axis: frequency in mHz, Y axis: value of $\gamma$ (pure number). Error bars have been magnified by a factor 50 in order to make them visible.}
  \label{fig:gamma}
\end{figure}

\begin{figure}[f]
  \centering
  \includegraphics[width=0.8\textwidth]{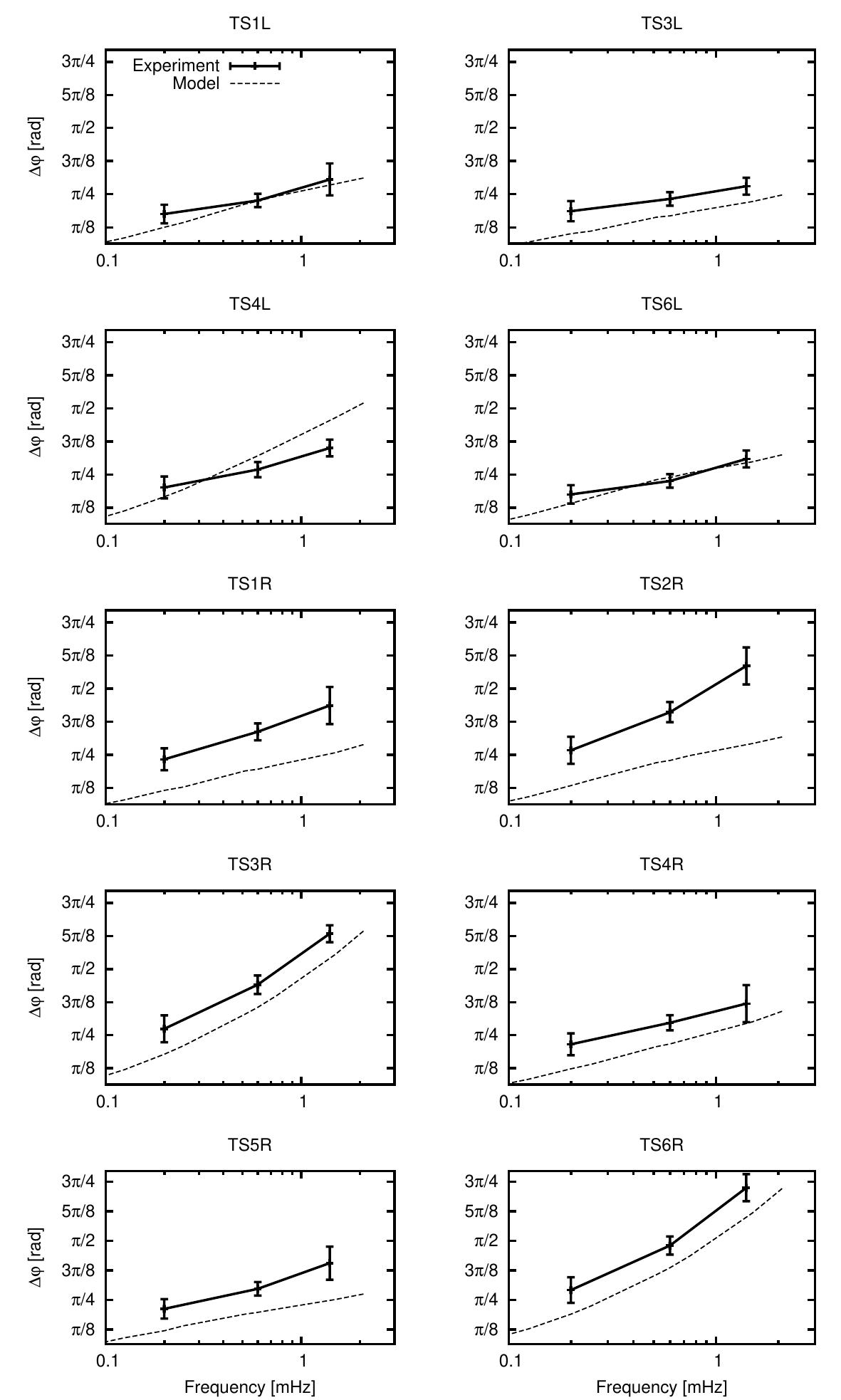}
  \caption{Comparison between the estimates of the numerical model and the experimental results for the $\Delta\varphi$ function (phase shift). X axis: frequency in mHz, Y axis: value of $\Delta\varphi$ (radians). Error bars have been magnified by a factor 10 in order to make them visible.}
  \label{fig:phi}
\end{figure}

\begin{table}[f]
  \caption{Estimated value of $\nu$, $\gamma$ and $\Delta\varphi$ for each of the three frequencies tested in the laboratory. This table only reports the best estimate among the ones obtained with the three methods (time-domain, Fourier, fitting) described in the text. The detailed results for $\gamma$ and $\Delta\varphi$ are reported in tables~\protect\ref{tbl:FMGamma} and \protect\ref{tbl:FMPhi}.}
  \centering
  \begin{tabular}{lr@{$\pm$}lr@{$\pm$}lr@{$\pm$}l}
    \hline
    Sensor   & \multicolumn{2}{c}{$\nu$ [mHz] } & \multicolumn{2}{c}{$\gamma$} & \multicolumn{2}{c}{$\Delta\varphi$ [rad]} \\
    \hline
   TS1L     &   0.1995 & 0.0008   &   0.4168 & 0.0008   &   0.551 & 0.011 \\
            &   0.5997 & 0.0009   &   0.2756 & 0.0004   &   0.711 & 0.008 \\
            &   1.3999 & 0.0008   &   0.2554 & 0.0003   &   0.960 & 0.019 \\
   \hline
   TS3L     &   0.1995 & 0.0008   &   0.5023 & 0.0010   &   0.584 & 0.012 \\
            &   0.5995 & 0.0012   &   0.3125 & 0.0005   &   0.729 & 0.008 \\
            &   1.4002 & 0.0008   &   0.2231 & 0.0003   &   0.880 & 0.010 \\
   \hline
   TS4L     &   0.1995 & 0.0008   &   0.4853 & 0.0010   &   0.632 & 0.013 \\
            &   0.5996 & 0.0010   &   0.2919 & 0.0004   &   0.844 & 0.009 \\
            &   1.4001 & 0.0011   &   0.1931 & 0.0002   &   1.100 & 0.010 \\
   \hline
   TS6L     &   0.1995 & 0.0008   &   0.5138 & 0.0010   &   0.550 & 0.011 \\
            &   0.5996 & 0.0010   &   0.3452 & 0.0005   &   0.708 & 0.008 \\
            &   1.4001 & 0.0004   &   0.2576 & 0.0003   &   0.970 & 0.010 \\
   \hline
   TS1R     &   0.1995 & 0.0008   &   0.5463 & 0.0011   &   0.733 & 0.013 \\
            &   0.5995 & 0.0013   &   0.2971 & 0.0006   &   1.059 & 0.010 \\
            &   1.3991 & 0.0025   &   0.1674 & 0.0002   &   1.371 & 0.022 \\
   \hline
   TS2R     &   0.1995 & 0.0008   &   0.5658 & 0.0011   &   0.839 & 0.016 \\
            &   0.5994 & 0.0015   &   0.3124 & 0.0007   &   1.292 & 0.012 \\
            &   1.3996 & 0.0017   &   0.1583 & 0.0002   &   1.841 & 0.022 \\
   \hline
   TS3R     &   0.1995 & 0.0008   &   0.5795 & 0.0012   &   0.860 & 0.016 \\
            &   0.5991 & 0.0021   &   0.2918 & 0.0009   &   1.385 & 0.011 \\
            &   1.4002 & 0.0016   &   0.1317 & 0.0002   &   1.990 & 0.010 \\
   \hline
   TS4R     &   0.1995 & 0.0008   &   0.4725 & 0.0009   &   0.677 & 0.013 \\
            &   0.5996 & 0.0011   &   0.2666 & 0.0005   &   0.933 & 0.009 \\
            &   1.3998 & 0.0011   &   0.1686 & 0.0002   &   1.160 & 0.022 \\
   \hline
   TS5R     &   0.1995 & 0.0008   &   0.4746 & 0.0010   &   0.668 & 0.013 \\
            &   0.5997 & 0.0009   &   0.3411 & 0.0005   &   0.934 & 0.009 \\
            &   1.3999 & 0.0008   &   0.2194 & 0.0003   &   1.276 & 0.022 \\
   \hline
   TS6R     &   0.1994 & 0.0008   &   0.4954 & 0.0009   &   0.918 & 0.017 \\
            &   0.5994 & 0.0014   &   0.2433 & 0.0005   &   1.510 & 0.012 \\
            &   1.3990 & 0.0026   &   0.1202 & 0.0002   &   2.277 & 0.018 \\
    \hline
  \end{tabular}
  \label{tbl:FMResults}
\end{table}

\begin{table}[f]
  \centering
  \caption{Estimated value of $\gamma$ for each of the three frequencies tested in the laboratory. All values are in radians. The results provided by each analysis method are reported here.}
  \begin{tabular}{lr@{$\pm$}lr@{$\pm$}lr@{$\pm$}l}
    \hline
    Sensor   & \multicolumn{2}{c}{Fourier} & \multicolumn{2}{c}{Time-domain} & 
    \multicolumn{2}{c}{Fit} \\
    \hline
    TS1L     &   0.417 & 0.002   &   0.416 & 0.004   &   0.4168 & 0.0008 \\
             &   0.275 & 0.002   &   0.273 & 0.003   &   0.2756 & 0.0004 \\
             &   0.256 & 0.004   &   0.257 & 0.003   &   0.2554 & 0.0003 \\
   \hline
    TS2L     &   0.527 & 0.002   &   0.526 & 0.004   &   0.5282 & 0.0008 \\
             &   0.432 & 0.001   &   0.433 & 0.003   &   0.4346 & 0.0005 \\
             &   0.478 & 0.004   &   0.477 & 0.004   &   0.4790 & 0.0005 \\
   \hline
    TS3L     &   0.502 & 0.003   &   0.500 & 0.005   &   0.5023 & 0.0010 \\
             &   0.311 & 0.002   &   0.312 & 0.004   &   0.3125 & 0.0005 \\
             &   0.222 & 0.004   &   0.225 & 0.004   &   0.2231 & 0.0003 \\
   \hline
    TS4L     &   0.486 & 0.003   &   0.483 & 0.005   &   0.4853 & 0.0010 \\
             &   0.291 & 0.002   &   0.293 & 0.003   &   0.2919 & 0.0004 \\
             &   0.193 & 0.004   &   0.192 & 0.004   &   0.1931 & 0.0002 \\
   \hline
    TS5L     &   0.536 & 0.002   &   0.538 & 0.003   &   0.5378 & 0.0008 \\
             &   0.448 & 0.001   &   0.451 & 0.002   &   0.4479 & 0.0005 \\
             &   0.532 & 0.004   &   0.528 & 0.003   &   0.5301 & 0.0006 \\
   \hline
    TS6L     &   0.515 & 0.003   &   0.507 & 0.004   &   0.5138 & 0.0010 \\
             &   0.344 & 0.002   &   0.345 & 0.003   &   0.3452 & 0.0005 \\
             &   0.259 & 0.004   &   0.262 & 0.002   &   0.2576 & 0.0003 \\
   \hline
    TS1R     &   0.547 & 0.003   &   0.545 & 0.005   &   0.5463 & 0.0011 \\
             &   0.298 & 0.003   &   0.294 & 0.004   &   0.2971 & 0.0006 \\
             &   0.168 & 0.004   &   0.171 & 0.003   &   0.1674 & 0.0002 \\
   \hline
    TS2R     &   0.569 & 0.004   &   0.564 & 0.006   &   0.5658 & 0.0011 \\
             &   0.313 & 0.003   &   0.313 & 0.005   &   0.3124 & 0.0007 \\
             &   0.158 & 0.003   &   0.159 & 0.004   &   0.1583 & 0.0002 \\
   \hline
    TS3R     &   0.582 & 0.004   &   0.576 & 0.006   &   0.5795 & 0.0012 \\
             &   0.293 & 0.003   &   0.291 & 0.004   &   0.2918 & 0.0009 \\
             &   0.130 & 0.003   &   0.131 & 0.004   &   0.1317 & 0.0002 \\
   \hline
    TS4R     &   0.474 & 0.003   &   0.471 & 0.003   &   0.4725 & 0.0009 \\
             &   0.264 & 0.002   &   0.265 & 0.002   &   0.2666 & 0.0005 \\
             &   0.169 & 0.003   &   0.170 & 0.002   &   0.1686 & 0.0002 \\
   \hline
    TS5R     &   0.476 & 0.003   &   0.473 & 0.003   &   0.4746 & 0.0010 \\
             &   0.340 & 0.003   &   0.340 & 0.003   &   0.3411 & 0.0005 \\
             &   0.219 & 0.004   &   0.222 & 0.002   &   0.2194 & 0.0003 \\
   \hline
    TS6R     &   0.498 & 0.003   &   0.493 & 0.003   &   0.4954 & 0.0009 \\
             &   0.246 & 0.002   &   0.243 & 0.003   &   0.2433 & 0.0005 \\
             &   0.119 & 0.002   &   0.118 & 0.003   &   0.1202 & 0.0002 \\
   \hline
  \end{tabular}
  \label{tbl:FMGamma}
\end{table}

\begin{table}[f]
  \centering
  \caption{Estimated value of $\Delta\varphi$ for each of the three frequencies tested in the laboratory. All values are in radians. The results provided by each analysis method are reported here. Note how the fitting method shows the largest discrepancies, although its error bars are always compatible with the other methods.}
  \begin{tabular}{lr@{$\pm$}lr@{$\pm$}lr@{$\pm$}l}
    \hline
    Sensor   & \multicolumn{2}{c}{Fourier} & \multicolumn{2}{c}{Time-domain} & 
    \multicolumn{2}{c}{Fit} \\
    \hline
    TS1L     &   0.551 & 0.011   &   0.548 & 0.042   &   0.73 & 0.18 \\
             &   0.711 & 0.008   &   0.715 & 0.028   &   0.61 & 0.20 \\
             &   0.960 & 0.019   &   0.974 & 0.044   &   0.94 & 0.03 \\
   \hline
    TS2L     &   0.315 & 0.006   &   0.306 & 0.034   &   0.47 & 0.15 \\
             &   0.311 & 0.004   &   0.309 & 0.023   &   0.28 & 0.12 \\
             &   0.438 & 0.011   &   0.440 & 0.034   &   0.43 & 0.01 \\
   \hline
    TS3L     &   0.584 & 0.012   &   0.577 & 0.044   &   0.77 & 0.19 \\
             &   0.729 & 0.008   &   0.732 & 0.032   &   0.57 & 0.25 \\
             &   0.885 & 0.018   &   0.879 & 0.048   &   0.88 & 0.01 \\
   \hline
    TS4L     &   0.632 & 0.013   &   0.617 & 0.045   &   0.83 & 0.20 \\
             &   0.844 & 0.009   &   0.856 & 0.034   &   0.73 & 0.21 \\
             &   1.129 & 0.021   &   1.114 & 0.052   &   1.10 & 0.01 \\
   \hline
    TS5L     &   0.301 & 0.006   &   0.298 & 0.026   &   0.46 & 0.16 \\
             &   0.290 & 0.004   &   0.295 & 0.020   &   0.27 & 0.12 \\
             &   0.428 & 0.010   &   0.432 & 0.032   &   0.42 & 0.02 \\
   \hline
    TS6L     &   0.550 & 0.011   &   0.544 & 0.032   &   0.73 & 0.18 \\
             &   0.708 & 0.008   &   0.708 & 0.024   &   0.60 & 0.21 \\
             &   0.983 & 0.019   &   0.990 & 0.042   &   0.97 & 0.01 \\
   \hline
    TS1R     &   0.733 & 0.013   &   0.727 & 0.039   &   0.93 & 0.20 \\
             &   1.059 & 0.010   &   1.059 & 0.033   &   0.89 & 0.26 \\
             &   1.371 & 0.022   &   1.376 & 0.059   &   1.27 & 0.09 \\
   \hline
    TS2R     &   0.839 & 0.016   &   0.828 & 0.045   &   1.05 & 0.21 \\
             &   1.292 & 0.012   &   1.301 & 0.036   &   1.09 & 0.30 \\
             &   1.841 & 0.022   &   1.802 & 0.067   &   1.76 & 0.05 \\
   \hline
    TS3R     &   0.860 & 0.016   &   0.849 & 0.043   &   1.08 & 0.22 \\
             &   1.385 & 0.011   &   1.389 & 0.035   &   1.03 & 0.44 \\
             &   1.982 & 0.024   &   1.988 & 0.075   &   1.99 & 0.01 \\
   \hline
    TS4R     &   0.677 & 0.013   &   0.660 & 0.031   &   0.87 & 0.19 \\
             &   0.933 & 0.009   &   0.930 & 0.026   &   0.79 & 0.24 \\
             &   1.160 & 0.022   &   1.178 & 0.044   &   1.13 & 0.03 \\
   \hline
    TS5R     &   0.668 & 0.013   &   0.660 & 0.030   &   0.87 & 0.20 \\
             &   0.934 & 0.009   &   0.936 & 0.025   &   0.85 & 0.18 \\
             &   1.276 & 0.022   &   1.260 & 0.042   &   1.25 & 0.03 \\
   \hline
    TS6R     &   0.918 & 0.017   &   0.903 & 0.033   &   1.14 & 0.21 \\
             &   1.510 & 0.012   &   1.523 & 0.029   &   1.32 & 0.29 \\
             &   2.277 & 0.018   &   2.271 & 0.060   &   2.18 & 0.10 \\
   \hline
  \end{tabular}
  \label{tbl:FMPhi}
\end{table}

\end{document}